\documentstyle{article}
\begin{document}
\title{\Large\bf Hall Resistivity in Ferromagnetic Manganese-Oxide Compounds \\}
\author{\large Liang-Jian Zou \\
    {\it  CCAST (World Laboratory) P.O.Box 8730, Beijing, 100080 } \\
    {\it  Institute of Solid State Physics, Academia Sinica,
          P.O.Box 1129, Hefei 230031, China$^{*}$}  \\}
\date{ }
\maketitle
\large
\noindent {\bf Abstract}  \\
 Temperature-dependence and magnetic field-dependence of the Hall effect and
the magnetic property in manganese-oxide thin films are studied. The
spontaneous magnetization and the Hall resistivity are obtained for a various
of magnetic fields over all the temperature. It is shown that the Hall
resistivity in small magnetic field is to
exhibit maximum near the Curie point, and strong magnetic field moves the
position of the Hall resistivity peak to much high temperature and suppresses
the peak value. The change of the Hall resistance in strong magnetic field may
be larger than that of the diagonal ones. The abnormal Hall resistivity
in the ferromagnetic manganese-oxide thin-films is
attributed to the spin-correlation fluctuation scattering.

\vspace{0.8cm}


\vspace{0.5cm}


\newpage
\noindent {\bf I. INTRODUCTION}

   Recently colossal negative magnetoresistance (MR) effect has been found in
ferromagnetic perovskite-like La$_{1-x}$R$_{x}$MnO$_{3}$ and
Nd$_{1-x}$R$_{x}$MnO$_{3}$ ( R =Ba, Sr, Ca, Pb, etc. )  [ 1 - 4 ]
thin films. In epitaxial La$_{1-x}$Ca$_{x}$MnO$_{3}$ and
Nd$_{1-x}$Sr$_{x}$MnO$_{3}$ thin films, the resistivity in applied magnetic
field decreases several order in magnitude under strong magnetic field,
which is much larger than that found in ferromagnetic/nonmagnetic metallic
multilayers. Since such a huge
MR change has potential application for magnetic
sensor, magnetic recording and many other aspects, it also presents another
possible MR mechanism which is different from that in magnetic multilayer,
it arises extensive attention and interests in the past two years [ 5 - 12 ].

Undoped La$MnO_{3}$ is an antiferromagnetic insulator [ 13 - 16 ], its magnetic
structure and magnetic properties had been studied in the early of 1950's.
After the substitution of a part of trivalent La or Nd ions by Ca, Sr, Pb,
Ba or other divalent elements,
Mn$^{+3}$ and Mn$^{+4}$ ions coexist, the valence fluctuation between them
is thus assumed to be important and may contribute to the hopping
conductivity and other transport properties. In the substitution
of La or Nd ions, La$_{1-x}$R$_{x}$MnO$_{3}$ and Nd$_{1-x}$R$_{x}$MnO$_{3}$
systems may occur the metal-insulator transition, the electronic states in
doped systems at the edge of metal-insulator transition can be heavily affected
by the long-range magnetic order and the external magnetic field. Some
mechanism have been proposed to explain the colossal MR effect in these
systems [ 8 - 11 ]. The magnetic polaron mechanism [ 1 - 3 ], the spin disorder
scattering [ 8, 9 ] and the metal-insulator transition [ 11 ]
induced by the magnetic field seem to be difficult to explain the colossal
MR behavior in strong magnetic field over all the temperature satisfactorily.

   In recent study [ 12 ], another possible mechanism for these ferromagnetic
La$_{1-x}$R$_{x}$MnO$_{3}$ and Nd$_{1-x}$R$_{x}$MnO$_{3}$ thin films is
proposed to explain the colossal MR. It is suggested that the abnormal large MR
in these compounds is induced by the spin-spin correlation fluctuation
scattering, the magnetic
property and the diagonal resistivity in different magnetic fields agree with
the experimental results very well. The Hall effect is one of the important
properties and may distinguish the present mechanism from the metal-insulator
transition mechanism [11], so it is useful to predict the Hall resistivity
behavior. In this letter, the Hall resistivity and the magnetic property
for temperature T and magnetic field B are obtained. The rest is arranged
as following, in Sec.II, the formalism is described; the results and
discussions are given in Sec.III, and the conclusion is drawn in Sec.IV.

\noindent {\bf II. FORMALISM}

   In La$_{1-x}$R$_{x}$MnO$_{3}$ and Nd$_{1-x}$R$_{x}$MnO$_{3}$ compounds,
because of the crystalline field effect, the 3d energy level of the Mn ion is
split into the low-energy triplet, t$_{2g}$, and the high-energy doublet,
e$_{g}$, therefore the three d-electrons of Mn$^{+3}$ and Mn$^{+4}$ ions
first fill the low t$_{2g}$ band, and the extra d-electrons in Mn$^{+3}$
will have to fill the high e$_{g}$ band, these two bands are separated
about 1.5 $eV$ [ 17 ]. Thus the three d-electrons in filled t$_{2g}$ band
forms a localized core spin through the strong Hund's coupling [ 13 - 16 ],
the core spins
tend to align parallelly through the double exchange interaction through
Mn$^{+3}$-O$^{-2}$-Mn$^{+4}$ and form ferromagnetic
background. The electrons in e$_{g}$ band are mobile and responsible for
electric conduction in these systems.
In this framework, the model Hamiltonian for the mobile
electrons moving in the ferromagnetic background is described:
\begin{equation}
       H=H_{0}+V
\end{equation}
\begin{equation}
   H_{0}=\sum_{k\sigma} (\epsilon_{k}-\sigma \mu_{B}B) c^{\dag}_{k \sigma}
         c_{k \sigma} -\sum_{<ij>} {\it A} {\bf S}_{i} \cdot {\bf S}_{j}-
         \sum_{i}g\mu_{B}BS^{z}_{i}
\end{equation}
\begin{equation}
     V = -\frac{J}{N} \sum_{ikq} \sum_{\mu \nu} e^{i{\bf q} {\bf R}_{i}}
       {\bf S}_{i} \cdot c^{\dag}_{k+q \mu} {\bf \sigma}_{\mu \nu} c_{k\nu}
\end{equation}
Where H$_{0}$ describes the bare energies of the mobile d-electrons and the
ferromagnetic background, V is the coupling
between the mobile electrons and the core spins. In Eq.(2), $\epsilon_{k}$
represents the energy spectrum of mobile (or conduction) electrons with respect
to the Fermi energy E$_{F}$; {\it A} denotes the effective ferromagnetic
exchange
constant between manganese ions, and only the nearest-neighbor interaction
is considered; -g$\mu_{B}$B is Zeemann energy in the magnetic field {\bf B}.
In Eq.(3),
the conduction electron is scattered from state k$\nu$ to state k+q$\mu$
by the localized spin ${\bf S_{i}}$; J denotes the
coupling between the conduction electrons and the core spins.
One notes that in the external magnetic field and the internal molecular
field of ferromagnetically ordered state, the conduction band
of the system will be split, this splitting is to move the position of the
conduction band with respect to the Fermi surface, thus the mean-field
spectrum of the conduction electron with state ${\bf k\sigma}$ is
$\epsilon_{k\sigma}=\epsilon_{k}-\sigma (\mu_{B}B+ J<S^{z}>)$.

    By the scattering theory [12], the lifetime of the conduction electrons
between two scatterings, $\tau$, is :
\begin{equation}
   \tau^{-1} = \frac{\pi}{h} \frac{J^{2}D(0)}{4} \sum_{kq\sigma}
             f_{k\sigma}(1-f_{k\sigma})f_{k+q\bar{\sigma}}(1-
             f_{k+q\bar{\sigma}})[<S^{-}_{q}S^{+}_{-q}>+
\end{equation}
\[   <S^{+}_{q}S^{-}_{-q}>+8 <S^{z}_{q}S^{z}_{-q}>] \]
Where D(0) is the density of states of the conduction electrons near the
Fermi surface, Dirac-Fermi distribution function
f$_{k}=1/[e^{\beta (\epsilon_{k}-\epsilon_{F})}+1]$. Then one can obtain the
diagonal conductivity through the Drude formula, $\rho_{xx}$=m/(ne$^{2}$$\tau$),
here n is the density of carrier concentration, and m the effective mass.
With the diagonal resistivity, one can easily derive the Hall conductivity:
in steady-state,
\begin{equation}
       \sigma_{H}=\frac{\sigma_{xx}^{2}}{ne} B_{eff}
\end{equation}
or the Hall resistivity:
\[    \rho_{H}=\frac{ne}{B_{eff}} \rho_{xx}^{2}  ~~~~~~~~~~~~~~~~~~~ (5')    \]
where B$_{eff}$ is effective field, B$_{eff}$=
$| {\bf B}+zA\ <{\bf S} \ >/\mu_{B}|$.

     From to Eqs.(4) and (5), one could qualitatively understand the
temperature-dependence of the Hall resistivity.
At zero-temperature limit (T $\rightarrow$ 0 K), $\rho_{xx}$ approaches
zero because of the Pauli exclusion principle, so $\rho_{H}$(T $\rightarrow$ 0)
$ \approx 0$; however at high temperature limit (T$\ >> T_{c}, E_{F}$),
$\rho_{xx}$ in different magnetic field approaches the same value, the Hall
resistivity mainly depends on the effective magnetic field. In this situation,
the diagonal and the Hall resistivities may exhibit some different behaviors.
In the mediated temperature range (0$<$k$_{B}$T  $<<$ E$_{F}$),
especially at the temperature near the Curie point, the diagonal resistivity
exhibits a maximum, therefore the Hall resistivity is also to exhibit a maximum,
however, because of the effect of magnetic field, the position of the maximum of
the Hall resistivity is smaller than that of the diagonal ones.\\

\noindent {\bf III. RESULTS AND DISCUSSIONS }

In quasi-2-dimensional systems, the lattice constant is a=3.89 $\AA$,
and the coordinate number is z=4. The theoretical parameters is from
La-Ca-Mn-O system, while the results can be applied for other similar systems.
In the calculation, several parameters need to be determined by the experiments
and the electronic structure calculation, for simplicity, the reduced resistivity
is adopted here.

    The dependence of the spontaneous magnetization and the Hall resistivity on
temperature is shown in Fig.1, for comparison, the diagonal resistivity is also
shown in Fig.1. The ferromagnetic-paramagnetic transition occurs within a
wide temperature range, because of the critical fluctuation and
the low-dimensional character, the temperature relation of the
spontaneous magnetization has a long tail, the transition temperature is
broaded significantly.
The electrons moving in the transverse direction is also scattered
by the spin-spin correlation fluctuation near the critical point,
therefore the Hall resistivity also exhibits a maximum. In the meantime,
the magnetic field {\bf B} affects the cyclotron movement of the
conduction electrons, the Hall resistivity decreases with {\bf B}, so
the maximum of the Hall resistivity will not appear at the same position
as that of the diagonal resistivity, the position of the maximum moves to
low temperature. It can be seen clearly in Fig.1.

   The field-dependence of the Hall
resistivity at different temperature is shown in Fig.2. One notes that
the Hall resistivity decreases monotonously when the external magnetic field is
increased. The descent
resistivity with the increase of magnetic field is due to the suppress
of the spin-spin correlation fluctuation scattering. Below T$_{c}$, the spin-spin
correlation is long range, the applied magnetic field has strong effect on the
scattering, so the change of the Hall resistivity is large (See the Curves (1)
and (2)); while above T$_{c}$,
the short-range spin-spin correlation fluctuation dominates the scattering of
conduction electrons, the external magnetic field has weak effect, so the
Hall resistivity exhibits weak dependence on the magnetic field.

   Since the transverse movement of the conduction electrons
are affected both by the magnetic field and by the spin-correlation
scattering, the change of the Hall resistivity may be larger than the
diagonal resistivity with the variation of the magnetic field. In the present
mechanism, from the curve (1) in Fig.2, the Hall resistivity may decrease two
to three orders in magnitude, for comparison, the diagonal resistivity only
changes one to two orders in magnitude.

\noindent {\bf IV. CONCLUSION}

   In conclusion, we have studied the Hall resistivity and the magnetic
property of the ferromagnetic manganese-oxide compounds. The
temperature-dependence of the Hall resistivity is to exhibit a maximum below
the peak position of the diagonal resistivity, and the change of Hall
resistivity in magnetic field will be larger than that of diagonal resistivity.
The spin-spin correlation scattering between the
conduction electrons and the localized spins is the intrinsic mechanism of
the abnormal Hall resistivity in La-R-Mn-O and Nd-R-Mn-O
compounds near the transition point.
 Further experiments is desired to verify our predictions.
 \\

   Acknowledgement: This work is financially supported by the Grant of
National Natural Science Funds of China No.19577101.

\newpage

\begin{center}
REFERENCES
\end{center}
~~~* mailing address
\begin{enumerate}
\item R. M. Kusters, J. Singleton, D. A. Keen, R. McGreevy and W. Hayes,
       {\it Physica} {\bf B155} 362 (1989).
\item R. Von Helmolt, J. Wecker, B. Holzapfeil, L. Schultz and K. Samwer
       {\it Phys. Rev. Lett.} {\bf 71} 2331 (1993).
      R. Von Helmolt, J. Wecker, and K. Samwer
       {\it J. Appl. Phys.} {\bf 76}, 6925 (1994).
\item S. Jin, T. H. Tiefel, M. McCormack, R. A. Fastnacht, R. Ramesh, and
      L, H, Chen  {\it Science}. {\bf 264} 413 (1994).
      {\it J. Appl. Phys.} {\bf 76}, 6929 (1994).
\item S. Jin, H. M. O'Bryan, T. H. Tiefel, M. McCormack, and W. W. Rhodos
        {\it Appl. Phys. Lett} {\bf 66}, 382 (1995).
\item G. C. Xiong, Q. Li, H. L. Ju, S. N. Mao, L. Senpati, X. X. Xi, R. L.
      Greene, and T. Venkatesan, {\it Appl. Phys. Lett} {\bf 66}, 1427 (1995).
\item H. L. Ju, C. Kwon, Qi Li, L.Greene and T. Vankateson,
       {\it Appl. Phys. lett.} {\bf 65}, 2106 (1994).
\item P. Schiffer, A. P. Ramirez, W. Bao and S-W. Cheong,
       {\it Phys. Rev.Lett.} {\bf 75}, 3336 (1995).
\item N. Furukawa, {\it J. Phys. Soc Jpn}, {\bf 63}, 3214 (1994).
\item J. Inoue and S. Maekawa, {\it Phys. Rev. Lett} {\bf 74}, 3407 (1995).
\item A. J. Millis , P. B. Littlewood, and B. I. Shraiman, {\it Phys.
      Rev. Lett} {\bf 74}, 3407 (1995).
\item A. Urushibara, Y. Moritomo, T. Arima, A. Asamitsu, G. Kido and Y. Tokuya,
      {\it Phys. Rev. B} {\bf 51}, 14103 (1995).
      Y. Moritomo, A. Asamitsu, and Y. Tokuya, {\it Phys. Rev. B} {\bf 51},
      16491 (1995).
\item Liang-Jian Zou, X. G. Gong, Q. Q. Zheng and C Y. Pan, {\it J. Appl. Phys.},
      {\bf 78}, No.4 (1996); {\it Phys. Rev. B}, submitted.
\item C. Zener, {\it Phys. Rev.}, {\bf 81}, 440 (1951); {\bf 82}, 403
      (1951).
\item E. O. Wollen and W. C. Koehler, {\it Phys. Rev.}, {\bf 100}, 545 (1955)
\item J. B. Goodenough, {\it Phys. Rev.}, {\bf 100}, 564 (1955)
\item P. G. De Gennes, {\it Phys. Rev.}, {\bf 118}, 141 (1960)
\item J. M. D. Coey, M. Viret and L. Ranno, {\it Phys. Rev. Lett.},
      {\bf 75}, 3910 (1995).

\end{enumerate}

\newpage
\begin{center}
Figures Captions
\end{center}

\noindent  Fig. 1.  Temperature-dependence of the spontaneous magnetization,
the Hall resistivity and the diagonal resistivity in magnetic field.
Theoretical parameters for calculation: {\it A}=145.5 K, {\it J}=350 K. B=15 T
(1) spontaneous magnetization, (2) Hall resistivity, and (3) diagonal resistivity.
\vspace{1cm}

\noindent  Fig. 2.  Dependence of the Hall
resistivity on the magnetic field at different temperature. Theoretical
parameters: {\it A}=145.5 K, {\it J}=350 K, (1) T=50 K, (2) T=100 K, (3) T=80 K.
\vspace{1cm}

\end{document}